\documentclass[a4paper,fleqn]{cas-dc}

\usepackage[numbers]{natbib}

\usepackage{subcaption}

\def\tsc#1{\csdef{#1}{\textsc{\lowercase{#1}}\xspace}}
\tsc{WGM}
\tsc{QE}

\begin{document}
\let\WriteBookmarks\relax
\def\floatpagepagefraction{1}
\def\textpagefraction{.001}
\shorttitle{Deep Learning for NEWSdm}
\shortauthors{A. Golovatiuk et~al.}

\title [mode = title]{Deep Learning for direct Dark Matter search with nuclear emulsions}                      



\author[1,2]{Artem Golovatiuk}[
                        orcid=0000-0002-7464-5675]
\cormark[1]
\ead{artem.golovatiuk@cern.ch}


\address[1]{I.N.F.N. sezione di Napoli, Napoli 80126, Italy}
\address[2]{Università degli Studi di Napoli Federico II, Napoli 80126, Italy}
\address[3]{National Research University Higher School of Economics, Moscow 101000, Russia}
\address[4]{National University of Science and Technology MISIS, Moscow 119049, Russia}
\address[5]{Lebedev Physical Institute of the Russian Academy of Sciences, Moscow 119991, Russia}

\author[3,4]{Andrey Ustyuzhanin}

\author[2,4,5]{Andrey Alexandrov}

\author[1,2]{Giovanni De Lellis}




\cortext[cor1]{Corresponding author}


\begin{abstract}
We propose a new method for the discrimination of sub-micron nuclear recoil tracks from instrumental background in fine-grain nuclear emulsions used in the directional dark matter search. The proposed method uses a 3D Convolutional Neural Network, whose parameters are optimised by Bayesian search. Unlike previous studies focused on extracting the directional information, we focus on the signal/background separation exploiting the polarisation dependence of the Localised Surface Plasmon Resonance phenomenon. Comparing the proposed method with the conventional cut-based approach shows a significant boost in the reduction factor for given signal efficiency.
\end{abstract}




\maketitle

\section{Introduction}
\label{sec:intro}

    Dark Matter (DM) is one of the remaining open questions of modern physics. While having numerous astrophysical indications of DM existence, we still have no evidence of the DM particles from ground-based experiments. Therefore it remains one of our priorities for a fundamental understanding of the structure of our Universe.
    
    Direct detection experiments are aiming at registering the signature of DM-nuclei scattering on the detector material. The typical range of DM particle masses and scattering cross-sections probed in these experiments corresponds to Weakly Interacting Massive Particles (WIMPs) \cite{wimp_dm, direct_dm, direct_wimp}. Within these experiments, some are trying to detect not only the event of the recoil and its energy but also the recoil direction. As we expect a flow of DM particles directed from the Cygnus constellation \cite{direct_dm, direct_astro}, the recoil direction gives us a possibility to better discriminate the detected signal from the isotropic background. 
    
    Since the expected DM interaction rate in the Standard Halo model~\cite{halo, halo-refine, direct_astro} is very low \cite{lewin}, every direct detection experiment needs to reduce as much as possible all sources of background contamination. Many background reduction techniques are applied at the design stage (such as underground location, shielding, material screening and purification). However, some level of background contamination remains, and additional algorithmic background reduction is needed.
    
    In this article, we present a Deep Learning (DL) based approach to analysing experimental data in the NEWSdm (Nuclear Emulsion for WIMP Search with directional measurement) experiment \cite{NEWSdm} and compare it with a more conventional cut-based approaches adopted by the NEWSdm. We repeat the analysis of an image brightness evolution due to a light polariser rotation in an optical microscope similar to \cite{super-res2020, super-res2019} and use the results of the analysis from \cite{NIT70}, where they use parameters of an optical image's elliptical fit to select signal-like candidates. The method in \cite{NIT70} looks only on ellipticity of the events without investigating a response to a polarised light (one unpolarised image per event).
    
    Our main goal is an effective reduction of a background contamination while keeping as much signal as possible. Experimentally, we can compensate a lower signal efficiency by collecting higher statistics. On the other hand, claiming a discovery in the search for very rare events requires extremely low levels of background. Additionally, we examine the influence of the compared classification approaches on the angular distributions.
    
    In sections  \ref{sec:news} and  \ref{sec:plasm} we describe the experimental setup and important physical effects used in the analysis. In sec.~\ref{sec:data} we provide details on the data, the details of the analysis approaches are in sec.~\ref{sec:analysis} and compare performances in sec.~\ref{sec:perform}.
    
    \subsection{The NEWSdm experiment}
    \label{sec:news}
    The NEWSdm experiment \cite{NEWSdm} uses silver bromide (AgBr) crystals dispersed in the gelatin medium both as a target material and as a tracking detector. The recoil nucleus leaves a track of activated crystals that are transformed into silver grains after chemical development and further remain stable. The particular type of emulsion used with fine crystal size of 75 nm and high granularity \cite{NIT, NIT70} allow registering tracks with lengths shorter than 100 nm, while preserving the directional information \cite{directionality} in tracks with at least two grains.
    
    The emulsion readout is performed with a custom-designed "super-resolution" optical microscope \cite{microscope, super-res2020}. This allows fast processing (e.g.~compared to an X-ray microscope) of the emulsion volumes and allows to scale up the detector mass.
    
    There remains some level of background contribution to the experimental tracks that needs to be analysed and discarded. Currently, two main background sources are intrinsic $\beta$ radiation from $^{14}C$ isotopes in gelatin and random thermal excitations of the crystals (also called fog) that convert into silver grains during chemical development. The background level due to the fog is expected to be about $10^{6}$ times higher than that from $\beta$ \cite{NEWSdm} and therefore is the most important to be suppressed.

    \begin{figure*}
     \centering
     \begin{subfigure}[b]{0.49\textwidth}
         \centering
         \includegraphics[width=\textwidth]{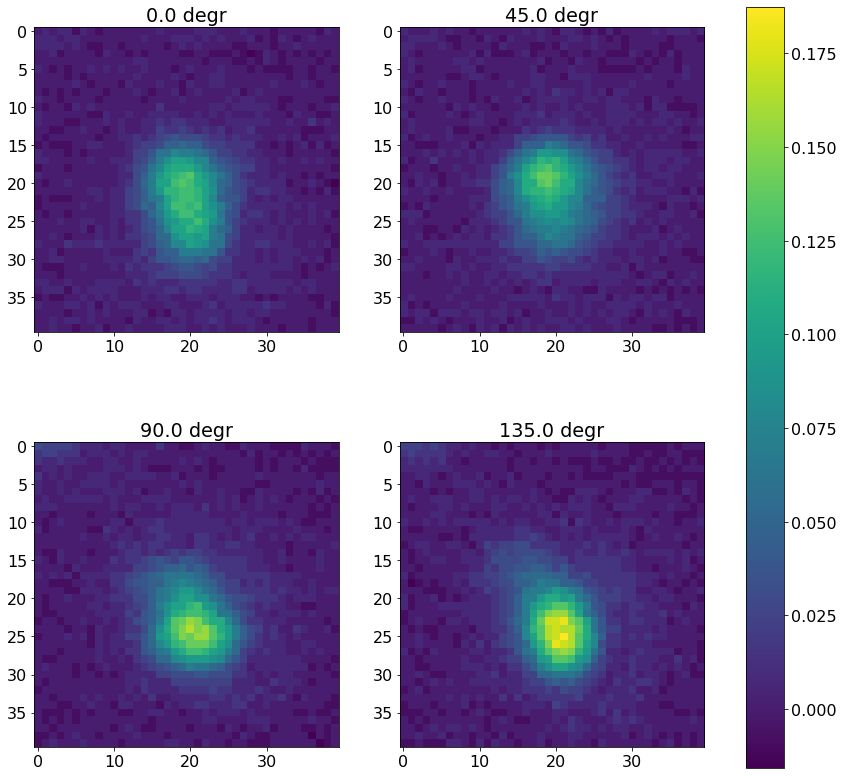}
         \caption{C100keV track}
         \label{fig:c100-im}
     \end{subfigure}
     \hfill
     \begin{subfigure}[b]{0.49\textwidth}
         \centering
         \includegraphics[width=\textwidth]{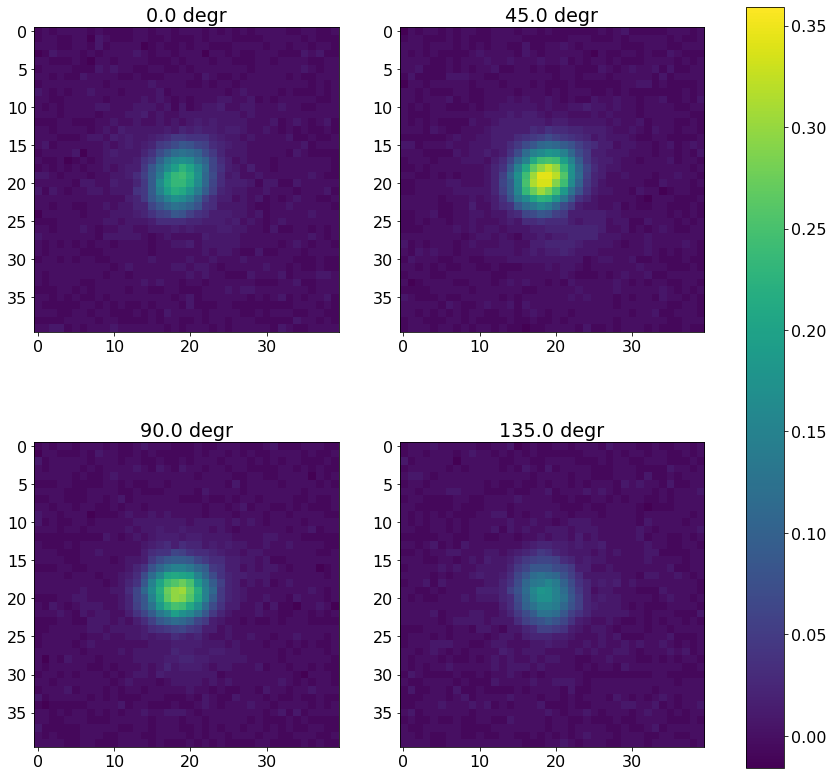}
         \caption{Fog grain}
         \label{fig:fog-im}
     \end{subfigure}
        \caption{Examples of polarised images for one signal event and one background (fog) event.} 
        \label{fig:pol-imgs}
    \end{figure*}
  
    \subsection{Plasmon resonance effect}
    \label{sec:plasm}
    The plasmon resonance effect comes from the interaction of linearly polarised light with small metallic objects \cite{microscope, super-res2020}. The intensity of reflected light depends on the polarisation angle of the illuminating light if the object has a non-spherical shape. If we approximate a silver grain by an ellipsoid, then we get a resonant peak at a longer light wavelength with polarisation along the major axis and a shorter wavelength along the minor axis. We observe the variation of the reflected intensity with the rotation of the polarisation angle for a light with a fixed wavelength.
    
    The silver grains have the form of randomly oriented filaments with the size of several tens of nanometers. Therefore, rotating the light polarisation produces resonant peaks at different angles for different grains. It causes the brightness peak of the image to shift between grains, producing a shift of the cluster barycenter. By studying the evolution of the image while rotating the polarisation direction, we can obtain additional insight into the internal structure of the track beyond the optical resolution. The barycenter shift is maximal when the grains' major axes are perpendicular and is close to zero when they are aligned. Therefore, it gives us a lower bound in estimating the real length of the corresponding tracks. The direction of the barycenter shift gives us the approximate angle of the original track's direction \cite{super-res2020}.

\section{Experimental data}
\label{sec:data}
    The data used in the analysis is currently purely experimental since Monte Carlo algorithms can not yet reproduce all the necessary physical processes (track formation, chemical development, polarised light interaction). Emulsion samples are either exposed to a specific source or unexposed to anything (random fog). The most significant background contribution is expected from thermal fog. Therefore, we study background reduction only on fog so far.
    
    The expected signal of nuclear recoils from WIMPs is experimentally imitated by beams of Carbon ions of fixed energies. A combination of samples exposed to Carbon can be used to simulate the recoil spectrum of WIMPs with the desired mass value. The beam energies used in this study are 30 keV and 100 keV. Due to the high intensity of Carbon tracks in the exposed areas, a tiny fraction of background contamination in these samples can be neglected.
    
    We have eight monochromatic images of $40\times40$ pixels from an optical microscope for each track, obtained at different polarisation angles, uniformly covering the $180^\circ$ period for a linear light polarisation. Independent areas of the same emulsions were used to develop the algorithms (training) and cross-check the final performance (testing). The validation data is a separate part of the "train" scans, used not for training but for selecting the optimal algorithm or model. Figure \ref{fig:pol-imgs} shows examples of microscope images taken at different polarisation angles for a Carbon 100 keV event (Fig.\ref{fig:c100-im}) and for a fog event (Fig.\ref{fig:fog-im}). The details on number of events for each class and dataset are presented in the Appendix~\ref{app:data}
    
    We apply some data cleaning prior to further analysis. We reject high event density areas, events with too few pixels, too low and too high brightness events, and non-isolated tracks (with multiple optically resolved peaks). The applied cuts reject most of the events caused by the dust particles, high-density regions (caused by physical imperfections in the emulsion such as scratches) and fake events when pixel noise is identified as an event with low brightness and just a few pixels. The same cleaning cuts are applied to all emulsion samples. This procedure is not expected to affect the tracks induced by WIMPs while reducing the background contamination.
    

    
    
    The Carbon 30keV emulsion sample was partially damaged on the surface (scratches and dust contamination), but we manually selected the cleaner areas to avoid imperfections that may contribute to fake signal events. 

    \subsection{Emulsion rotations}
    \label{sec:emul-rot}
    Due to vibrations of the microscope during the data acquisition, the measured barycenter shifts have some anisotropy since images for different polariser rotations are taken sequentially. It is not expected to bring problems in background scanning since the latter is isotropic. However, the Carbon samples are exposed in a specific direction, and therefore measurement anisotropy can interfere with it.
    
    We manually rotate the emulsions and use a randomly shuffled mixture of events from the rotated scans in the DL training to make our analysis independent of the correlation between the beam direction and measurement anisotropy. We scan 4 independent areas for each Carbon sample: first with the default orientation ($0^\circ$ rotation), second rotated by $45^\circ$, third rotated by $90^\circ$ and again with the default orientation check reproducibility of the results. 
    
    \begin{figure}[htbp]
    \centering 
    \includegraphics[width=.35\textwidth]{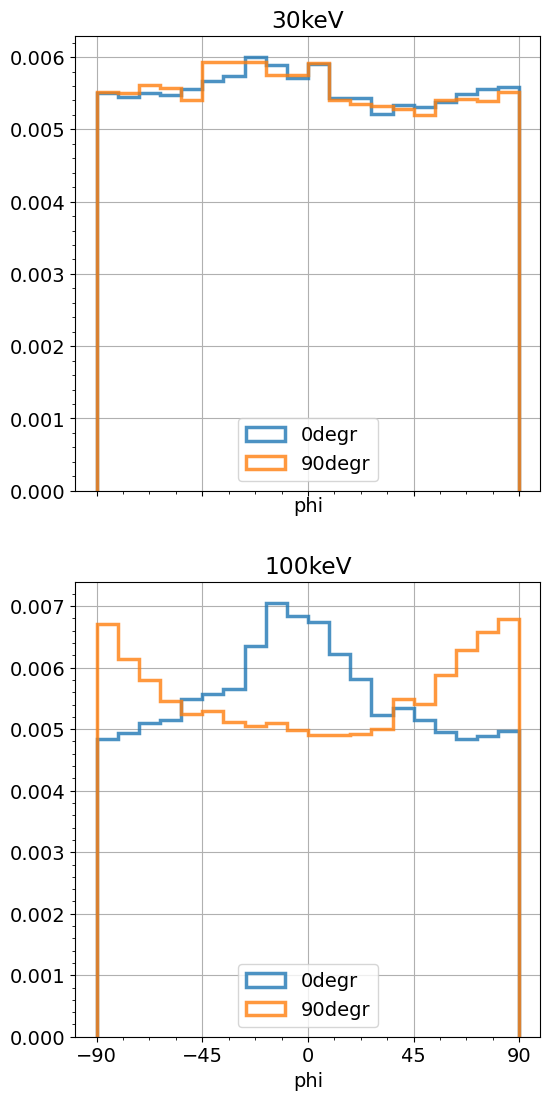}
    \caption{\label{fig:angle_rot} Comparison of the track angular distributions in rotated samples for different Carbon beam energies. Each histogram is normalised so the area under the histogram integrates to 1.}
    \end{figure}
    
    The Carbon beams were along $0^\circ$ in the default orientation, therefore the physically relevant angular peak is expected in the same direction. Comparing the angular distributions estimated similar to \cite{super-res2020} from the image's barycenter movement due to the variation of the polarisation angle for samples scanned in the default orientation and rotated by $90^\circ$ (Fig.\ref{fig:angle_rot}), one can see that measurement anisotropy dominates the directional information in the lower energy 30 keV sample, while we see the clear directional peaks in 100 keV sample. However, this does not mean that no directional information is present in the 30 keV sample.  The lower beam energy produces shorter tracks and the scatter worsens the reported angular resolution.  Since the barycenter shift measurement is not isotropic (independent of phi), the small and wide peak can be hidden by the shape of the anisotropy with this method.

\section{Analysis and methods}
\label{sec:analysis}
    In the following we briefly describe the conventional analysis approach followed by a detailed description of the proposed DL-based one. Both approaches use the same microscope images taken at several polarisation angles.

    \subsection{Conventional barycenter shift analysis}
    
    
    \begin{figure}[htbp]
    \centering 
    \includegraphics[width=.35\textwidth]{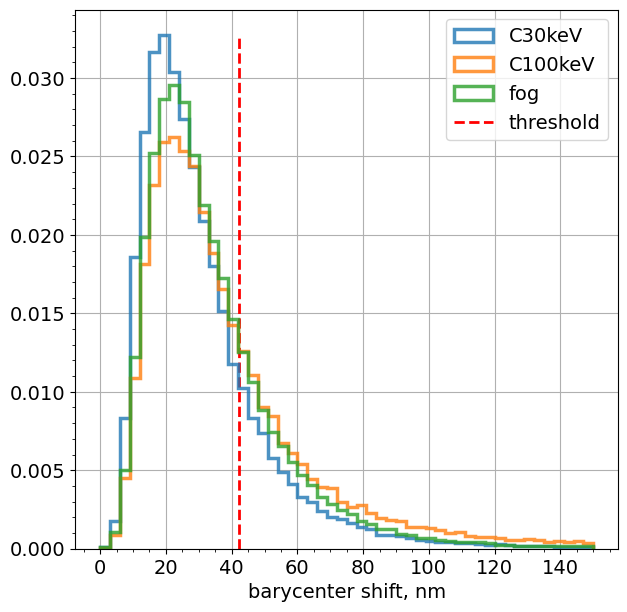}
    \caption{\label{fig:bar_thresh} Barycenter shift distributions for tracks reconstructed in emulsions exposed to Carbon beams and unexposed. The unexposed sample is representative of the random fog. The threshold for short-long split is shown by the red dashed line.}
    \end{figure}
    
    The barycenter shift produced by the variation of the polarisation angle can be used to try to distinguish single grains and tracks composed of more than one grain. The higher energy recoils tend to produce longer tracks \cite{directionality}. We expect background to be mostly represented by single-grain events, while signal events need to be multi-grain to measure the direction. Therefore, we can use a threshold on barshift to split events into "long" and "short" tracks, and classify "short" track events as background. We calculate the barycenter shift similar to the \cite{super-res2020, super-res2019}.
    
    Figure \ref{fig:bar_thresh} shows barycenter shift histograms for Carbon and fog samples with the threshold fixed to provide same signal efficiency of $30\%$ for Carbon 100keV on validation data as the method in \cite{NIT70}. Carbon 100keV has the highest beam energy among the samples in the analysis, and thus better approximates the long-track signal. The threshold is set to $42.3\pm 0.2$ nm.
    
    
    

    \subsection{DL-based analysis}
    We study a machine learning approach to signal-background classification with Convolutional Neural Networks (CNNs) \cite{CNN, CNN_intro}. The main goal is the strongest achievable background reduction factor while keeping efficiency (percentage of signal that passes the selection) similar to the conventional analysis. The study is performed in the Python language \cite{python} using the Keras library \cite{keras} for building and using the network.
    
    The network takes as an event input a 3D numeric array (2D polarised images stacked together to provide an additional dimension) and outputs 3 probability-like numbers (in the range from 0 to 1 which sum up to 1). The event outputs can be treated as pseudo-probabilities for corresponding classes: Carbon 30 keV, 100 keV and fog.
    
    A realistic representation of the WIMP signal needs a combination of Carbon samples with different energies (approximating a continuous spectrum), so we want our network to identify all Carbon as a signal but preserve some knowledge about different energies. Although, we have a limited set of Carbon beam energies available in this study, the approach is generalisable to a wider dataset. Therefore, we define the task as a multi-class problem with a mixed loss function \eqref{balanced-loss}. 
    
    \begin{equation}
        \label{balanced-loss} L_{\text{balanced}} = L_{\text{multi}} + \mu_{l} L_{\text{binary}}
    \end{equation}
    
    One part of the loss is categorical cross-entropy, defined in eq.\eqref{cross-entr}, another is a binary cross-entropy (the same concept, but for two classes), that is multiplied by a hyper-parameter $\mu_{l}$ with a value optimised by the Bayesian search described in sec.\ref{sec:bayes}. The three classes in eq.\eqref{cross-entr} are Carbon 30 keV, 100 keV and fog, while for the binary case, we join all Carbons into a `signal' class against `background' (represented only by fog).
    
    \begin{equation}
        \label{cross-entr} L_{\text{multi}} = \sum_i y_i \log (\bar{y_i})
    \end{equation}
    where $\bar{y_i}$ is the vector of the output predictions of different classes for the specific event and $y_i$ is the vector of the correct answers, but with zeros corresponding to all classes except one.
    
    To obtain a discrete answer from the classifier, we sum outputs corresponding to Carbon classes and obtain a scalar probability of being signal for an event. Then need to select a threshold on this probabilistic output. We choose a threshold value that provides the same signal efficiency for Carbon 100keV as in barshift and shape analysis \cite{NIT70}. This way we can compare the improvement in background reduction.
    
    \subsubsection{Image preprocessing}
    We try to achieve several goals by preprocessing the images passed to the network: improving the convergence during training, reducing the impact of the instrumental noise, and insert some physical features into the data.
    
    \begin{itemize}
        \item \textbf{Image scaling.} For training convergence purposes, we scale images by the maximum pixel value (defined by the camera specifications) before further processing. This value is the same for all images to avoid distortions by scaling.
        
        \item \textbf{Optical background subtraction.} The optical background around the track can contain distinguishable features not related to the event itself. Due to differences in production processes, Carbon-exposed and Fog samples have different thicknesses and event densities leading to slightly different image background brightness due to different properties of diffused light.
    
        To prevent algorithm from learning from the background brightness alone, we calculate the median brightness of each image and subtract the median value of the specific image from all its pixels (independently for every polarisation). Hence, the optical background brightness values are always centred around zero with slight variations (the camera noise amplitude is much lower than the scaling value). As a consequence, the contrast (signal-to-noise ratio) in the images is improved as well.
        
        \item \textbf{Periodic boundary conditions.} The polarisation angle has a period of $180^\circ$. If we add another rotation after the last ($180^\circ$), we will get the same polariser orientation as the first image ($0^\circ$). In order to account for this effect in our network, we add a copy of the first image after the last one for each event, getting nine images in total.
        
        \item \textbf{Random rotations.} The directionality of the signal is an important feature, which we want to preserve in DL analysis. To prevent the remaining background (false positives after classification) from adopting the same angular distribution as the signal, we apply algorithmic random rotations to every event. All nine images of the same track are rotated by the same angle to preserve correlations between polarisations. The pixels of a rotated image are interpolated and points outside the boundaries are padded with a reflection of nearby values.
    \end{itemize}
    
    \subsubsection{Convolutional Neural Network}
    We are using the architecture inspired by ResNet \cite{resnet, resnet2} and optimised for our specific task. The NEWSnet network we developed consists of basic building blocks like convolution layers~\cite{CNN_intro}, SWISH activation functions~\cite{swish}, batch normalisation (BN)~\cite{batch-norm}, max-pooling~\cite{CNN_intro} and dropout layers~\cite{CNN_intro}. The architecture is made of convolutions and groups of residual skip connections (identity shortcuts and convolution shortcuts as defined in \cite{resnet2}). The output is obtained from the softmax activation function with the corresponding number of classes (2 Carbon types and 1 fog), while for binary loss and final background reduction performance, the probabilities of signal classes are summed. We are using $3\times3\times3$ filters within convolution layers. The receptive field of the filter grows for every next layer, so this size is enough to catch all the correlations in the data. The usage of larger filters is not justified as it would explode the number of trainable network parameters without obvious benefits. Residual connections also contain layers with $1\times1\times1$ filters, as suggested in \cite{resnet, resnet2}. The details for every layer are provided in Appendix \ref{app:network}. 
    
    As we have important physical correlations stored in images for different polarisation angles, we consider a set of 9 images as one 3D image. We use 3D versions for all the CNN components. This allows the network to better find physical correlations along the "polarisation axis". The images are monochromatic, hence they have only one colour channel. The resulting architecture is general-purpose and the network is fitted to a specific problem during training.
    
    A detailed description of the NEWSnet architecture together with hyper-parameter values are provided in the Appendix \ref{app:network}. A schematic view of a similar network architecture was provided in \cite{acat_poster}. The number of skip-connections, number of filters in the layers, regularisation parameters (dropout rates) and training parameters (such as learning rate) are optimised with a Bayesian search \cite{bayes}.
    
    

    \subsubsection{Bayesian search}
    \label{sec:bayes}
    We use a Bayesian optimisation strategy to search for the optimal hyper-parameters \cite{bayes}. The Bayesian approach is an improved version of random search over the parameter space, where the probability distribution from which the parameters are sampled, is not uniform but is updated using Gaussian Processes~\cite{gaussproc} in order to optimise the goal function: area under the receiver operating characteristic (ROC) curve (AUC ROC) for binary classification in our case. We use AUC ROC to select the best parameter set independently from other methods.
    
    We implement this algorithm using the Scikit-optimize package \cite{scikit-opt}. The hyperparameters being optimised are:
    \begin{itemize}
        \item The dropout rates (3 dropout layers: before the residual blocks, in between and after final residual block).
        \item Number of convolution filters in the layers.
        \item Number of residual identity shortcuts after each residual convolutional shortcut \cite{resnet2}.
        \item Proportion multiplier $\mu_{loss}$ for the combined loss \eqref{balanced-loss}.
        \item Learning rate (multiplier of the gradient step).
        \item $\beta_1$ and $\beta_2$ parameters of Adam optimizer \cite{adam} (decay rates for the 1st and 2nd momenta)
        \item Exponential decay rate for the learning rate.
    \end{itemize}
    
    For each set of parameters, we trained on half of the data and for fewer epochs to speed up the search, with 3-fold cross-validation, to obtain more robust results for the goal function. We take the parameter set corresponding to the best result on validation after 150 iterations of Bayesian search. The exact values and the constraints within which the search was performed are presented in the Appendix \ref{app:network}.
    
\section{Performance comparison}
\label{sec:perform}
    
    We compare both methods discussed in sec.~\ref{sec:analysis} with the results reported in \cite{NIT70} for the same emulsion type, but obtained with ellipticity cuts for optical images in unpolarised light. Each approach has its benefits and limitations, which we discuss in this section.
    
    \subsection{Barycenter shift thresholds}
    \label{sec:barshift-perform}
    
\begin{figure}
    \centering 
    \includegraphics[width=.48\textwidth]{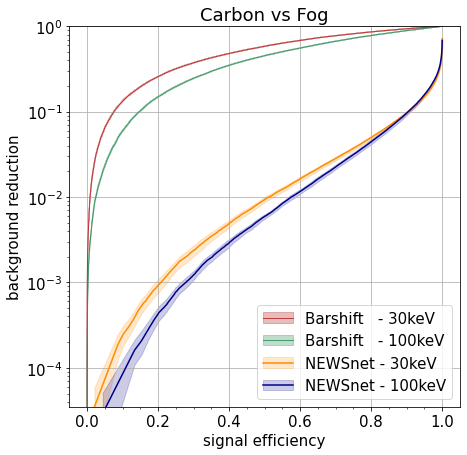}
    \caption{\label{fig:ml_bar_roc} Signal efficiency against background reduction for different signal classes (Carbon beam energies) against Fog background.}
    \end{figure}

    The barycenter shifts on fig.\ref{fig:bar_thresh} follow the physical intuition for Carbon exposed samples, having in general longer track events for higher beam energies. However, it does not allow to separate them from the fog background effectively. The percentage of events surpassing the threshold (Tab.~\ref{tab:perform}) is comparable for Carbon and fog samples and does not provide a way to effectively reduce the background, as it reduces the signal by the same amount. The illustrative comparison of background reduction for a given signal efficiency with the NEWSnet is plotted on Fig. \ref{fig:ml_bar_roc}.
    
\begin{figure*}
    \centering 
    \includegraphics[width=.95\textwidth]{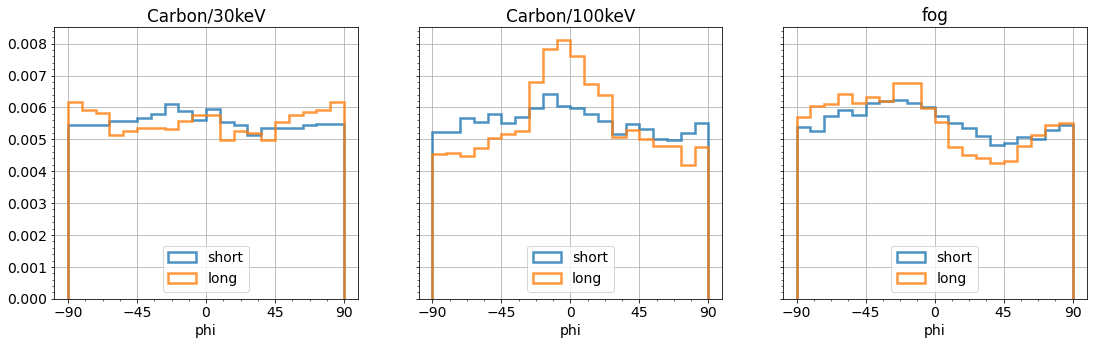}
    \caption{\label{fig:angle_thresh} Angular distributions for the events with the barycenter shift longer and shorter than the threshold in different emulsion samples. Each histogram is normalised independently to integrate to 1.}
    \end{figure*}

    Figure \ref{fig:angle_thresh} compares directions of the short and long track events. 100 keV Carbon events have a clear directional peak in the long tracks, while short follow the anisotropy pattern. 30 keV Carbon does not produce a clear direction as physical peak is expected at $0^\circ$. The small peak at $90^\circ$ for long tracks was not expected and is a subject for a future cross-check whether it is caused by a sample contamination. The anisotropy pattern is more significant in long tracks for the fog events, as this anisotropy can to contribute to elongating the single-grain events by slightly shifting the image between polariser rotations.
    
    \subsection{Deep Learning selection}
    We visualise the performance as a ROC curve: the horizontal axis is True Positive rate (signal efficiency), the vertical axis is False Positive rate (the background reduction factor). We took the model configuration with the highest AUC ROC on cross-validation in the Bayesian search (sec.\ref{sec:bayes}) and trained it on the entire training set for 60 epochs. We used networks from the last ten epochs of the training to obtain an ensemble to compute mean performance and confidence intervals on the results. The networks are coming from the same training to assess the behaviour of this specific trained network for future applications. The performance on the validation set is presented on Fig.~\ref{fig:ml_bar_roc}.
    
    To obtain a binary classifier, we select a threshold on the probabilistic output of the network on the validation set and cross-check the resulting performance on the test set (separate scannings of the independent areas of the same emulsions). We fix a threshold of 0.9975 and present the results in Table \ref{tab:perform}. This threshold allows us to achieve $\sim 10^{-3}$ background reduction while preserving a significant part of the signal. We choose the threshold to achieve same $30\%$ signal efficiency for Carbon 100keV as other methods in the comparison.
    
    \begin{table*}[width=1.8\linewidth,cols=6,pos=htbp]
    \caption{Comparison of the signal efficiencies and background reduction factors between barycenter shift analysis and NEWSnet on validation and test data. Last column contains the efficiencies of the shape analysis \cite{NIT70} performed on the same emulsion type.}\label{tab:perform}
     \begin{tabular*}{\tblwidth}{@{} L *{5}{C} @{} }
     \toprule
      & \multicolumn{2}{c}{Barshift} & \multicolumn{2}{c}{NEWSnet} & Shape analysis \\ 
     & Validation & Test & Validation & Test & \\
     \midrule
      \multicolumn{6}{c}{Signal efficiency} \\
     \midrule
     C30keV & $18.9 \pm 0.4\%$ & $19.3 \pm 0.4\%$ & $23.1 \pm 3.7\%$ & $12.2 \pm 2.8\%$ & $1.7\pm 0.1\%$\\
     C100keV & $30.0 \pm 0.4\%$ & $30.3 \pm 0.5\%$ & $30.0 \pm 3.4\%$ & $31.1 \pm 3.3\%$ & $29.7 \pm 0.7\%$ \\
     \midrule
     \multicolumn{6}{c}{Background reduction factor} \\
     \midrule
     Fog & $0.25 \pm 0.01$ & $0.25\pm0.01$ & $(1.4\pm0.45)\cdot 10^{-3}$ & $(2.7\pm0.81)\cdot 10^{-4}$ & $0.01$ \\
     \bottomrule
    \end{tabular*}
    \end{table*}
    
    The contamination in the Carbon 30keV sample (mentioned in sec.~\ref{sec:data}) has a different pattern in different areas. Thus, the independent test area yields lower network output values, and the signal efficiency is decreased in this sample. Similarly, the network output is lower on the Fog test events, which gives us stronger reduction factor. In general, this is acceptable since our primary goal is to reject anything that we suspect is not a signal. 
    
    The NEWSnet provides significant improvement in performance comparing to the shape analysis reported in the last column of the Table~\ref{tab:perform}. The background studied in \cite{NIT70} is represented by a silver nanoparticle sample with high event density in order to simulate chance coincidences of multiple spherical particles located closer than the optical resolution. In that study they consider it similar to the reference fog sample that also represents random coincidences of single grains. More precise comparison can be achieved if both approaches are applied to the same emulsion sample.
    
\begin{figure*}
     \centering
     \begin{subfigure}[b]{0.95\textwidth}
         \centering
         \includegraphics[width=\textwidth]{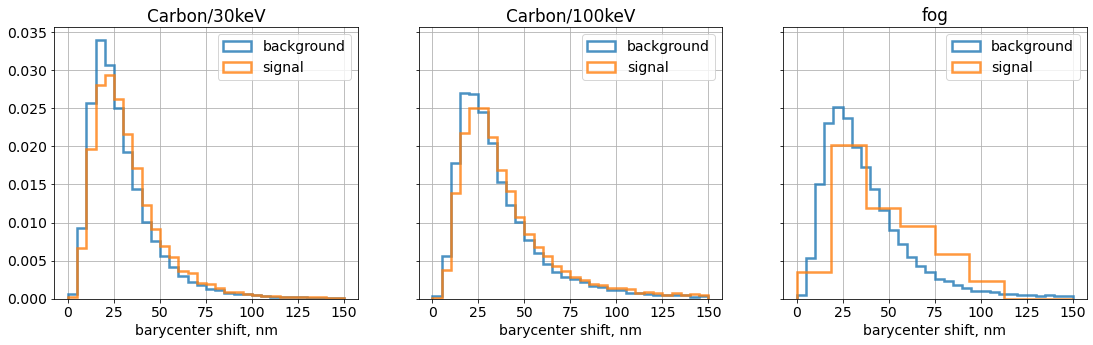}
         \caption{Barycenter shifts}
         \label{fig:ml_barshift}
     \end{subfigure}
     \hfill
     \begin{subfigure}[b]{0.95\textwidth}
         \centering
         \includegraphics[width=\textwidth]{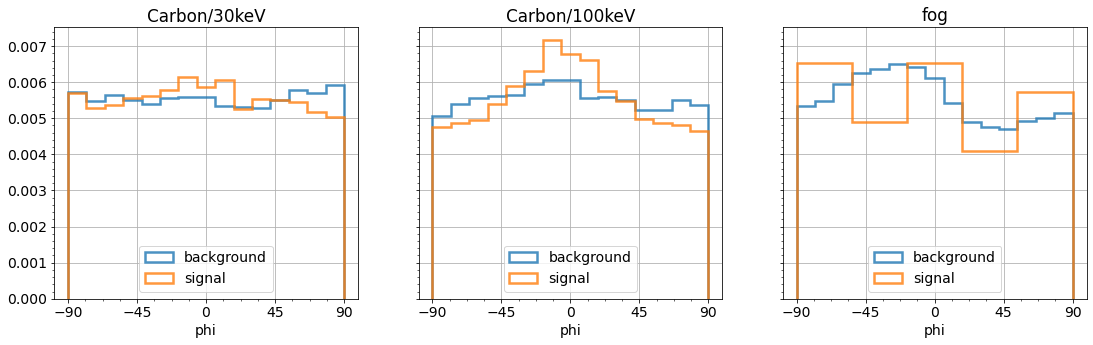}
         \caption{Angular distributions}
         \label{fig:ml_phi}
     \end{subfigure}
        \caption{Physical features for the events selected as signal or background by the NEWSnet in different emulsion samples. Each histogram is normalised independently to integrate to 1.} 
        \label{fig:ml-selected}
    \end{figure*}

    We checked the same physical features of polarised images as in sec.~\ref{sec:barshift-perform} for the classification obtained by DL. Figure \ref{fig:ml-selected} show the barycenter shift and angular distributions for classified events. The "signal" distributions for the fog sample are less accurate due to the low amount of fog selected as a signal in test data (48 events).
    
    The "signal" barycenter shifts are, in general, slightly longer but we do not observe a significant difference on fig.~\ref{fig:ml_barshift}. Therefore, DL selection does not use barycenter shift as a distinguishing feature, which is expected since barshift was not successfull at distinguishing background from signal. Angular distributions in fig.~\ref{fig:ml_phi} show directional peak only in "signal" events in Carbon 100keV sample, similarly to fig.~\ref{fig:angle_thresh}. The "signal" events in the fog sample do not tend to have a directional peak. However, a much bigger dataset is needed to get a more precise distribution.
    
    The signal efficiencies of both barycenter shift analysis and the NEWSnet are higher than the ones with shape analysis. However, we are able to identify the directional peak in the signal only for 100 keV Carbon. While the shape analysis detects signal direction in the 30 keV Carbon, the efficiency is very low and comparable to the remaining background fraction. All these approaches need further improvements to identify direction and keep a larger fraction of signal for low recoil energies, as they correspond to smaller WIMP masses.
    
\section{Conclusions}
\label{sec:concl}
    In the presented study, we compared conventional cut-based and DL-based approaches for background reduction in a nuclear emulsion WIMP search experiment. We used images from an optical microscope with a rotating polarising filter for the candidate events. 
    
    The first approach is based on the displacement of the brightness peak due to resonant effects of polarised light reflected from non-spherical silver grains. It estimates track length lower bound and direction. It does not provide an efficient way of signal-background separation based on barycenter shift threshold. However, it still obtains a clear directional peak in preselected events for Carbon 100keV beam.
    
    The second approach is DL-based. We use 3D CNN to predict the track class. The final background reduction factor is around $10^{-3}$, and it is stronger on the independent data. By fixing the signal efficiency of the final model similar to the cut-based approaches, we obtained orders of magnitude higher background reduction. 
    
    The events selected as a signal by the network show the directional peak only in the Carbon 100keV sample, which is similar to the cut-based barycenter shift analysis. 
    
    Due to the data preprocessing steps we apply, the DL selection does not imply the directional information in the selected data. Therefore this physical feature in the selected candidates can be considered intact and used for further analyses.
    
    
    The NEWSnet can provide good background reduction and preselect a reduced number of candidates for further detailed study while preserving intact the directionality. 

The reported network has replaced the cut-based analysis in the NEWSdm experiment in order to improve the background reduction and the sensitivity to WIMPs. The extended version of the network, including electron background samples, is under development and will be used for the analysis of the 10 kg emulsion detector, proposed by the NEWSdm collaboration. We expect that addition of new physical variables, like the plasmon wavelength information, to the input of the network can boost the background reduction orders of magnitude thus pushing the NEWSdm emulsion detector’s sensitivity towards the levels required to reach the neutrino floor where it will be able to benefit from the use of the directionality information.

\section*{Declaration of competing interest}

The authors declare that they have no known competing financial interests or personal relationships that could have appeared to influence the work reported in this paper.

\section*{Acknowledgements}

This work is supported by a Marie Sklodowska-Curie Innovative Training Network Fellowship of the European Commissions Horizon 2020 Programme under contract number 765710 INSIGHTS.

The research leading to these results has received funding from Russian Science Foundation under grant agreement n$^{\circ}$ 17-72-20127.

\appendix

\section{The NEWSnet architecture and parameters}
\label{app:network}
    The network starts with 3D convolutions, followed by the residual blocks defined in table \ref{tab:res-blocks}, max pooling and dropout layers, finishing with a fully connected layer with softmax activation to produce the output predictions of different classes. The architecture with configurations for all the layers is presented in table \ref{tab:newsnet}. There was an additional third dropout layer during the Bayes optimisation before the final fully-connected one, but the optimal dropout rate for it turned out to be 0. The batch size is 256.
    
    The parameter constrains and optimal values:
    
    \begin{itemize}
        \item The 3 dropout layers: before the residual blocks in range [0.1,0.4], the final value is 0.4; in the middle in range [0.3,0.8], the final value is 0.3; after the final residual block in range [0,0.4], the final value is 0.
        \item Number of convolution filters in the layers: for the residual blocks before the second dropout layer in range [32,64], the final value is 64; for the residual blocks after the second dropout layer in range [64, 128], the final value is 128.
        \item Number of residual identity shortcuts after each residual convolutional shortcut in range [0,3], the final value is 3.
        \item Proportion multiplier $\mu_{loss}$ for the combined loss \eqref{balanced-loss} in range [0.01, 2], the final value is 0.53.
        \item Learning rate in range $[10^{-5},3\times10^{-3}]$, the final value is $6.5\times10^{-4}$.
        \item Parameters of Adam optimizer \cite{adam}: $\beta_1$ in range [10, 100], the final value is 10; $\beta_2$ in range [100, 1500], the final value is 100.
        \item Exponential decay rate for the learning rate in range $[10^{-4}, 10^{-1}]$, the final value is $10^{-1}$.
    \end{itemize}
    
    The total amount of trainable parameters in the final architecture is 6 925 910.
    
    \begin{table*}[width=1.6\linewidth,cols=2,pos=hbtp]
    \caption{Residual building blocks with (N,N,4N) filters. Strides are 1 when not specified otherwise.}\label{tab:res-blocks}
     \begin{tabular*}{\tblwidth}{@{} L L @{} }
     \toprule
      \multicolumn{2}{c}{Residual block: identity shortcut \cite{resnet}} \\ 
     \midrule
     $1\times1\times1$ 3D Conv., N filters, BN, SWISH & \\
     $3\times3\times3$ 3D Conv., N filters, BN, SWISH & Identity path \\
     $1\times1\times1$ 3D Conv., 4N filters, BN & \\
     \midrule
     \multicolumn{2}{c}{Addition of paths} \\
     \multicolumn{2}{c}{SWISH} \\
     \bottomrule
     \toprule
      \multicolumn{2}{c}{Residual block: convolution shortcut \cite{resnet2}} \\
     \midrule
     $1\times1\times1$ 3D Conv., 2 stride, N filters, BN, SWISH & \\
     $3\times3\times3$ 3D Conv., N filters, BN, SWISH & $3\times3\times3$ 3D Conv., 2 stride, 4N filters, BN \\
     $1\times1\times1$ 3D Conv., 4N filters, BN & \\
     \midrule
     \multicolumn{2}{c}{Addition of paths} \\
     \multicolumn{2}{c}{SWISH} \\
     \bottomrule
    \end{tabular*}
    \end{table*}

    \begin{table}[width=.9\linewidth,cols=1,pos=h]
    \caption{Final NEWSnet architecture.}\label{tab:newsnet}
     \begin{tabular*}{\tblwidth}{@{} L @{} }
     \toprule
     $3\times3\times3$ 3D Conv., 32 filters, BN, SWISH \\
     $3\times3\times3$ 3D Conv., 64 filters, BN, SWISH \\
     MaxPooling 3D \\
     Dropout 0.4 rate \\
     Residual convolution block (64,64,256) filters \\
     Residual identity block (64,64,256) filters \\
     Residual identity block (64,64,256) filters \\
     Residual identity block (64,64,256) filters \\
     MaxPooling 3D \\
     Dropout 0.3 rate \\
     Residual convolution block (128,128,512) filters \\
     Residual identity block (128,128,512) filters \\
     Residual identity block (128,128,512) filters \\
     Residual identity block (128,128,512) filters \\
     Flatten \\
     FC layer 3 units \\
     Softmax \\
     \bottomrule
    \end{tabular*}
    \end{table}
    

\section{Dataset sizes and details}
\label{app:data}
    The data used in the study is represented by sets of nine optical images of 40x40 pixel for each event. The emulsion scans with manual rotations mentioned in sec.~\ref{sec:emul-rot} are joined together (equal parts for every rotation) and randomly split into the Training and Validation sets. The Test set is produced by independent scannings without emulsion rotations. The number of events used in the study are presented in Table \ref{tab:data}.
    
    \begin{table}[width=.9\linewidth,cols=4,pos=h]
    \caption{Dataset composition.}\label{tab:data}
     \begin{tabular*}{\tblwidth}{@{} L *{3}{C} @{} }
     \toprule
     & Train & Validation & Test \\
     \midrule
     C30keV & 39000 & 26000 & 50000\\
     C100keV & 39000 & 26000 & 50000 \\
     Fog & 51300 & 34200 & 50000 \\
     \bottomrule
    \end{tabular*}
    \end{table}

\newpage


\bibliographystyle{model1a-num-names}

\bibliography{cas-refs}





\end{document}